\documentclass[paper]{agujournal2019}
\usepackage{url}
\usepackage{lineno}
\usepackage{color}
\draftfalse

\newcommand{\planss} {{Planatary Space Science }}  
\newcommand{\ssr}{   {Space Sci. Rev. }}

\newcommand{\jgr}{   {J. Geophys. Res.}}
\newcommand{\grl}{   {Geophys. Res. Lett.}}
\newcommand{\apj}{   {Astrophys. J.}}
\newcommand{\apjl}{   {Astrophys. J. Lett.}}

\newcommand{\nat}{   {Nature}}
\newcommand{\icarus}{   {Icarus}}
\newcommand{\aap}{   {Astronomy \& Astrophysics}}

\journalname{JGR: Space Physics}

\begin{document}


\title{Force-free current sheets in the Jovian magnetodisk: the key role of electron field-aligned anisotropy}

\authors{A. V. Artemyev \affil{1}, Q. Ma\affil{2,3}, R. W. Ebert \affil{4,5}, X.-J. Zhang\affil{6,1},  F. Allegrini\affil{4,5}}
\affiliation{1}{Department of Earth, Planetary, and Space Sciences, University of California, Los Angeles, USA}
\affiliation{2}{Department of Atmospheric and Oceanic Sciences, University of California, Los Angeles, CA, USA}
\affiliation{3}{Center for Space Physics, Boston University, Boston, MA, USA}
\affiliation{4}{Southwest Research Institute, San Antonio, TX, USA}
\affiliation{5}{Department of Physics and Astronomy, University of Texas at San Antonio, San Antonio, TX, USA}
\affiliation{6}{Department of Physics, University of Texas at Dallas, Richardson, TX, USA}

\correspondingauthor{Anton Artemyev}{aartemyev@igpp.ucla.edu}

\begin{keypoints}
\item We report Juno observations of thin anisotropic current sheets in the Jovian magnetodisk
\item The contribution of electron streams to the current sheet stress balance is estimated
\item We show force-free current sheet configuration supported by strong electron field-aligned currents
\end{keypoints}

\begin{abstract}
Current sheets are an essential element of the planetary magnetotails, where strong plasma currents self-consistently support magnetic field gradients. The current sheet configuration is determined by plasma populations that contribute to the current density. The most commonly investigated configuration is supported by diamagnetic cross-field currents of hot ions, typical for the magnetospheres of magnetized planets. In this study, we examine a new type of the current sheet configuration supported by field-aligned currents from electron streams in the Jovian magnetodisk. Such bi-directional streams increase the electron thermal anisotropy close to the fire-hose instability threshold and lead to strong magnetic field shear. The current sheet configuration supported by electron streams is nearly force-free, with $|{\bf B}|\approx const$ across the sheet. Using Juno plasma and magnetic field measurements, we investigate the internal structure of such current sheets and discuss possible mechanisms for their formation.
\end{abstract}

\section{Introduction}
\begin{figure*}
\centering
\includegraphics[width=1\textwidth]{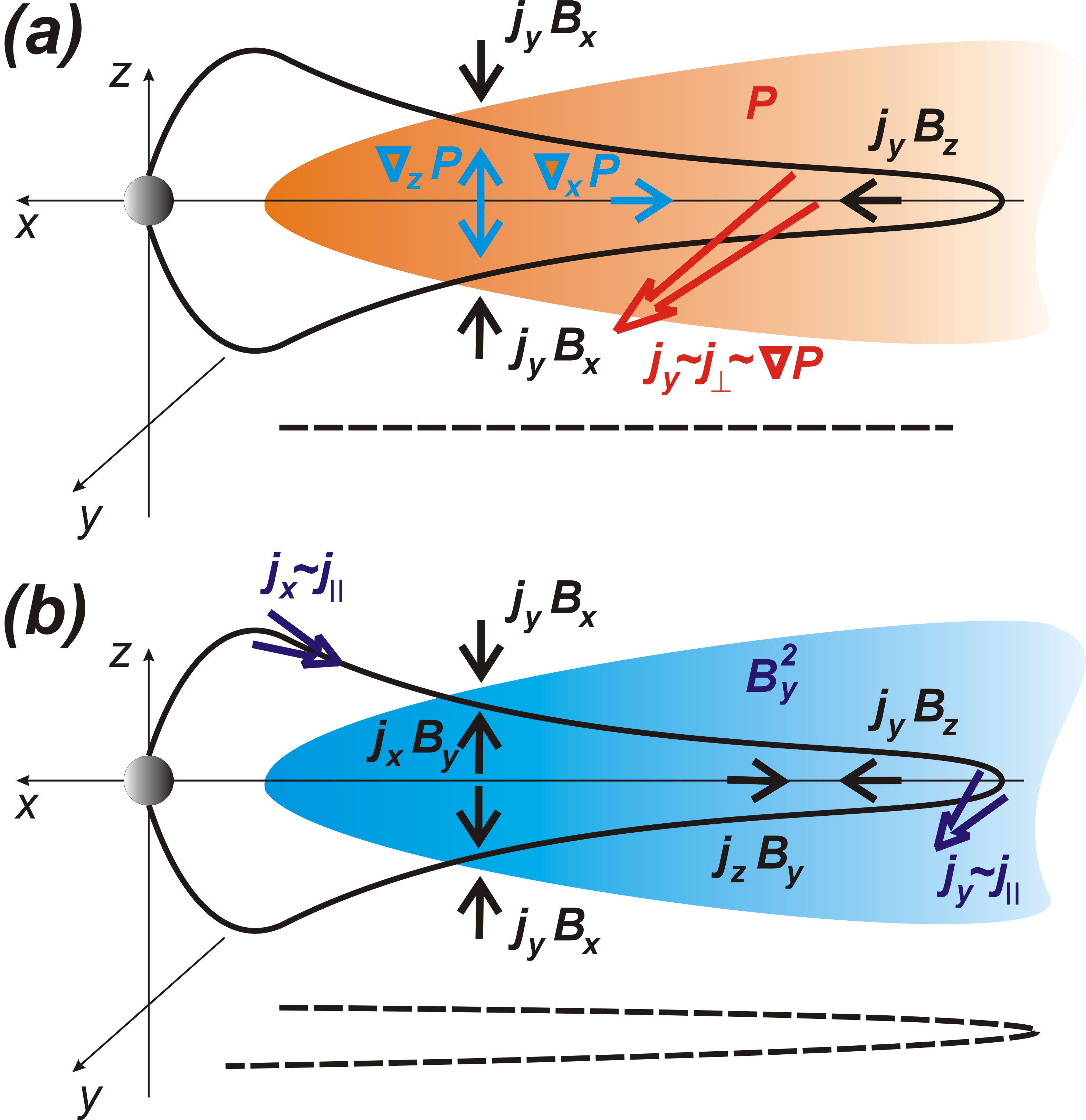}
\caption{Schematic of current sheet configurations. (a) 2D configuration of the magnetic field line in the current sheet with plasma pressure $P$ gradients (the dashed line parallel to the $x$-axis illustrates its projection to the $xy$ plane). The main current density flows along the $y$-axis, transversely to magnetic field lines. (b) 3D configuration of the magnetic field line in the current sheet with $B_y^2/8\pi$ gradients playing the role of $P$ gradients \cite<see discussion on the $P\to B_y^2$ transition in>{Syrovatskii81,Lukin18}. The dashed curve illustrates the field line projection to the $xy$ plane. The main current density flows along ${\bf B}$, i.e., parallel to the magnetic field line.}
\label{fig1}
\end{figure*}

Current sheets are observed in all planetary magnetotails, the night-side regions of stretched magnetic field lines. The configuration of magnetotails depends on characteristics of the planetary magnetic field interaction with solar wind \cite{Bagenal92,bookBagenal00,Jackman14,Khurana&Liu18}, but all magnetotails contain current sheets, spatially localized regions of strong plasma currents. Instabilities of such current sheets, either internally or externally driven, can result in the magnetic reconnection that further transforms the magnetic energy to the plasma heating and charged particle acceleration \cite<e.g.,>{Birn12:SSR,book:Gonzalez&Parker,Sitnov19}. Among all the current sheets in space plasmas, the one in Earth's magnetotail has been most intensively investigated, where strong diamagnetic currents, predominantly carried by hot protons (with a small fraction of oxygen ions), support the magnetic field configuration and pressure balance self-consistently (see Schematic in Fig. \ref{fig1}(a) and \citeA{Birn04,Sitnov&Merkin16,Zelenyi11PPR,Artemyev&Zelenyi13,Lu16:CS}). The current sheet in Earth's magnetotail is characterized by large plasma $\beta\sim 100$ ($\beta$ is the ratio of plasma and magnetic field pressures), which leads to the dominant role of cross-field currents, $j_\perp\gg j_\parallel$, in the current sheet configuration. The relative contribution of ions and electrons to $j_\perp$ depends on the polarization electric fields, ${\bf E}_\perp$ \cite{SB02,Schindler12}; thin (ion-kinetic-scale) current sheets are, therefore, mostly electron dominated \cite{Hesse98,Runov06,Artemyev09:angeo,Lu19:jgr:cs} due to the strong polarization electric field within \cite{Zelenyi10GRL,Lu16:CS}. However, in the Earth's magnetotail, there are two interesting exceptions that $j_\parallel$ can be appreciably large. First, in the near-Earth magnetotail, during the current sheet thinning (formation of thin current sheets during the substrom growth phase, see \citeA{Birn04MHD,Petrukovich07,Hsieh&Otto15}) the spatial scale (thickness) of the current sheet can become smaller than the thermal proton gyroradius. In such sub-ion scale thin current sheets the proton pressure cannot be redistributed within sub-gyroradius scale. To establish the pressure balance, the intensified field-aligned electron currents form strong magnetic field shear, contributing to a (partially) force-free current sheet configuration with $j_\perp \leq j_\parallel$  (see Schematic in Fig. \ref{fig1}(b) and \citeA{Nakamura08,Artemyev13:jgr,Artemyev17:grl:currents}). Second, a similar, partially force-free current sheet configuration has been observed in the distant (lunar orbit) magnetotail \cite{Xu18:artemis_cs} where plasma $\beta$ can be as low as $\sim 1$, and plasma pressure is not sufficient to establish the pressure balance. This second force-free current sheet with low $\beta$ is also typical in Mars \cite{Artemyev17:jgr:mars,DiBraccio15} and Venus \cite{Rong15:venus} magnetotails occupied by cold planetary plasmas.

Between the two mechanisms for the force-free (or partially force-free) current sheet formation, sub-ion scale thickness versus low plasma $\beta$, the first one is more interesting, because such sub-ion current sheets with strong field-aligned electron currents can be favorable to kinetically-driven magnetic field reconnection and current filamentation \cite{Drake&Lee77,Zelenyi&Taktakishvili87,Wilson16:cs}. Investigations of such current sheet configurations, however, are quite limited, because these current sheets are rather transient (dynamical) in the Earth’s magnetotail \cite<see discussions in>{Nakamura08}. An alternative plasma environment to investigate these specific current sheets would require hot heavy ions, curved magnetic field lines, and fast electron plasma flows. The best possible, accessible system is the Jovian magnetodisk, filled by sulfur and oxygen ions \cite{Thomas04:inbook, Krupp04, Mauk04,Kim20:juno} with various charge states \cite<e.g.,>{Selesnick09, Clark16, Allen19,Kim20:juno:PS} and conjugate to the Jovian aroural region, a powerful source of field-aligned electron streams \cite<e.g.,>{Mauk17,Mauk17:nature}. Therefore, We will use the recently available plasma and magnetic field measurements \cite{Bagenal17} from Juno in the Jovian magnetodisk to systematically examine sub-ion scale, force-free (or partially force-free) current sheets.

In this study, we focus on 18 events of Juno current sheet crossings during the first 30 orbits, i.e., those with a strong magnetic shear (field-aligned currents), electron field-aligned streams, and different combinations of proton and heavy ion contributions to the pressure. We estimate the current sheet spatial scale (thickness) and current density during its flapping motion. The following of paper includes three sections: description of Juno instruments and data analysis techniques in Sect.~\ref{sec:data}, detailed analysis of 9 current sheet crossings in Sect.~\ref{sec:cs}, and discussion on the results in Sect.~\ref{sec:discussion}.

\section{Data analysis technique and instruments}\label{sec:data}
\begin{figure*}
\centering
\includegraphics[width=1\textwidth]{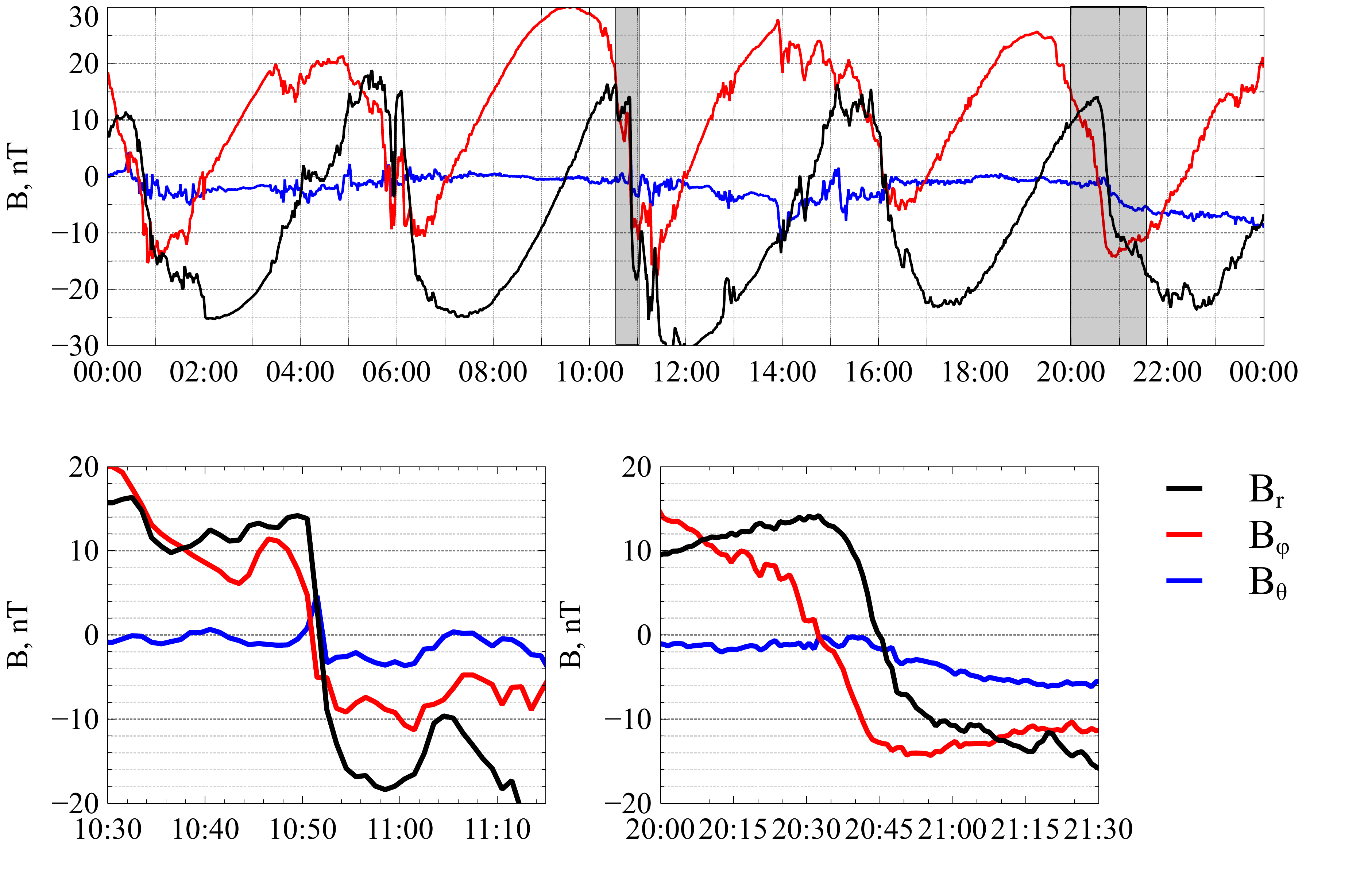}
\caption{Top panel shows one-day measurements by Juno MAG: radial $B_r$, azimuthal $B_\varphi$, and north-south $B_\theta$ components. Bottom panels show the expanded view of two current sheet crossings from the top panel.}
\label{fig2}
\end{figure*}

We use data from the Juno magnetometer (MAG) in 2017-2018, with $1$s time resolution \cite{Connerney17:ssr, Connerney17:science}. We focus on measurements at $r>25R_J$ radial distances in the Jovian magnetodisk. Figure \ref{fig2} shows a typical one-day magnetic field measurements in the magnetodisk by Juno: quasi-periodic crossings of zeros of the radial magnetic field component $B_{r}=0$ (current sheet) are due to flapping oscillations of the magnetodisk. For each such crossing we transform the magnetic field into local coordinate systems \cite{Sonnerup68}: $B_l$ is the most varying magnetic field component, $B_n$ is the less varying component, and $B_m$ is transverse to $B_l$ and $B_n$. We keep only those current sheets with a $B_m$ peak at $B_l=0$ and with available plasma measurements by the Jovian Auroral Distributions Experiment (JADE) instrument. The times of selected crossings are given in table \ref{table}, along with their radial distances. Overview plots of plasma and magnetic field profiles during each current sheet crossing are provided in the Supporting Information. In the main text, we mostly discuss six force-free current sheets, in comparison with three current sheets supported by plasma pressure gradients (non-force-free sheets), but our conclusions are supported by the entire dataset.

Jovian Auroral Distributions Experiment (JADE) measures \cite<see>{McComas17:ssr,McComas17,Kim20:juno,Kim20:juno:PS} electron distributions from below $\sim0.1$ to $100$ keV (at a $1$ s cadence) and ions from $\sim13$ eV to $\sim50$ keV (including ion composition, at a $1$ s cadence). We have averaged JADE data over the spin period ($30$ s) to obtain a complete pitch angle coverage from $0^{\circ}$ to $180^{\circ}$. These energy ranges cover the main (thermal) plasma populations in Jupiter's magnetodisk. We use the following data products from JADE: electron pitch-angle/energy distributions averaged over time interval of the current sheet crossing, electron omnidirectional energy spectra (energy flux) $F_e(t,E)$, electron pressure $p_e(t)$ and density $n_e(t)$ profiles, electron pressure anisotropy $A_e=p_{e,\parallel}/p_{e,\perp}$ averaged over the current sheet crossing interval, omnidirectional proton and heavy ion energy spectrum (energy flux) $F_p(t,E)$ and $F_i(t,E)$, and proton and heavy ion densities $n_p(t)$, $n_i(t)$. Note that field view of the JADE electron detector  may not cover the entire $[0,180^\circ]$ pitch-angle range at the current sheet boundaries, so we average electron measurements over each current sheet crossing interval to obtain a more complete pitch-angle distribution. The rotation of the background magnetic field direction across the current sheet ensures a wide coverage of the pitch-angle range (see pitch-angle distributions below), which is needed for the estimate of $A_e$. In this study, we use the {\it heavy ion} (or "i") to denote integrated quantities of those ion populations with mass-to-charge ratio larger than five. We also estimate proton and heavy ion pressures $p_p(t)$, $p_i(t)$ as moments of the omnidirectional energy spectrum, i.e., we assume that thermal proton and ion speeds well exceed their bulk flow speed \cite<this is a reasonable assumption, see>{Kane99,Frank02Jupiter,Waldrop05,Kim20:juno:PS}.

\begin{table}
\centering
\begin{tabular}{ c|c|c|c|c|  }
\# & date & time & radial  & comments\\
&  &  & distance & \\
 \hline
 1 & 2017 doy 027 & 09:00-09:20 & $65R_J$ & $ff$ CS,  heavy ions \\
 2 & 2017 doy 080 & 07:10-07:40 & $61R_J$ & $ff$ CS,  heavy ions \\
 3 & 2017 doy 080 & 17:10-17:50 & $61R_J$ & non-$ff$ CS,  heavy ions \\
 4 & 2017 doy 128 & 08:00-08:30 & $86R_J$ & $ff$ CS,  heavy ions \\
 5 & 2017 doy 128 & 16:30-17:30 & $86R_J$ & $ff$ CS,  heavy ions \\
 6 & 2017 doy 133 & 05:10-06:10 & $61R_J$ & non-$ff$ CS,  heavy ions \\
 7 & 2017 doy 181 & 06:50-07:20 & $86R_J$ & $ff$ CS,  heavy ions \\
 8 & 2017 doy 181 & 09:45-10:20 & $86R_J$ & $ff$ CS,  heavy ions \\
 9 & 2018 doy 031 & 09:30-10:40 & $50R_J$ & non-$ff$ CS,  heavy ions \\
10 & 2018 doy 034 & 13:30-14:30 & $50R_J$ & non-$ff$ CS,  heavy ions \\
11 & 2018 doy 085 & 21:00-21:50 & $62R_J$ & non-$ff$ CS,  protons \\
12 & 2018 doy 086 & 10:15-11:15 & $55R_J$ & $ff$ CS,  heavy ions \& protons \\
13 & 2018 doy 088 & 01:20-02:30 & $38R_J$ & $ff$ CS,   protons \\
14 & 2018 doy 088 & 11:30-12:30 & $38R_J$ & $ff$ CS,  protons \\
15 & 2018 doy 088 & 21:30-22:30 & $38R_J$ & $ff$ CS,  heavy ions  \\
16 & 2018 doy 141 & 10:30-11:15 & $38R_J$ & $ff$ CS,  heavy ions \& protons  \\
17 & 2018 doy 141 & 19:45-21:00 & $38R_J$ & $ff$ CS,  heavy ions \& protons  \\
18 & 2018 doy 142 & 20:00-20:45 & $26R_J$ & non-$ff$ CS,  heavy ions \& protons  \\
\hline
\end{tabular}
\caption{List of current sheet crossings. In the comments column, $ff$ CS and non-$ff$ CS stand for the force-free and non force-free current sheets, respectively. The dominant ion type (heavy ions or protons) is also indicated in the comments column. \label{table}}
\end{table}

Using plasma and magnetic field measurements, we estimate $\beta_{e,p,i}=8\pi p_{e,p,i}/B^2$ profiles and electron fire-hose parameter $\Lambda_e\approx \beta_{e}(A_e-1)/2=4\pi (p_{\parallel e}-p_{\perp e})/B^2$. To show that this parameter controls the contribution of the electron anisotropy to the current density, we illustrate the case for a simple quasi-1D current sheet with $\partial B_l/\partial r_n=4\pi j_m/c$, $\partial B_m/\partial r_n = 4\pi j_l/c$, and $B_n=const\ne 0$. The electron current due to cross-field drifts in such current sheet is \cite{BookSJB66}
\begin{equation}
 {\bf j}_ {\bot e}   =  - ecn\frac{{\left[ {{\bf E} \times {\bf B}} \right]}}{{B^2 }} - c\frac{{\left[ {\nabla  p_ {\bot e}   \times {\bf B}} \right]}}{{B^2 }} + \frac{c\Lambda_e}{4\pi} \frac{{\left[ {{\bf B} \times \left( {{\bf B}\nabla } \right){\bf B}} \right]}}{{B^2 }}  \label{eq:currents_per}
\end{equation}
where ${\bf E}_\perp$ is the transverse component of the polarization electric field. This equation should be supplemented by the field-aligned stress balance equation
\[
enE_\parallel  =  - \nabla _\parallel p_{\parallel e}  +\frac{\Lambda_e}{4\pi}\left( {{\bf B}\nabla } \right)B
\]
where $E_\parallel$ is the field-aligned component of the polarization electric field and $\nabla_\parallel=({\bf B}/B)\nabla$.

For force-free current sheets with $\nabla_n p_{e\bot}=\nabla_n p_{e\parallel}=\nabla_n B=0$, the current density equation can be rewritten as
\[
{\bf j}_ {\bot e}   = \frac{c\Lambda_e}{4\pi} \left( { - {\bf e}_l \frac{{B_n^2 }}{{B^2 }}\nabla _n B_m  + {\bf e}_m \frac{{B_n^2 }}{{B^2 }}\nabla _n B_l  + {\bf e}_n \frac{{B_n }}{{B^2 }}\left( {B_l \nabla _n B_m  - B_m \nabla _n B_l } \right)} \right)
\]
whereas the parallel current density can be obtained from the divergence free condition:
\[
j_{\parallel e}  =  - \frac{1}{B}\left( {B_l \nabla _n B_m  - B_m \nabla _n B_l } \right)
\]
Thus, the total electron current is
\begin{eqnarray}
 {\bf j}_e  &=& {\bf j}_{ \bot e}  + j_{\parallel e} \frac{{\bf B}}{B} = 4\pi c\Lambda _e \left( { - {\bf e}_l \left( {\nabla _n B_m  - \frac{{B_m \nabla _n B}}{B}} \right) + {\bf e}_m \left( {\nabla _n B_l  - \frac{{B_m \nabla _n B}}{B}} \right)} \right) \nonumber\\
 \label{eq:currents_par}\\
  &=& \frac{c\Lambda _e}{4\pi} \left( { - {\bf e}_l \nabla _n B_m  + {\bf e}_m \nabla _n B_l } \right) = \Lambda_e {\bf j} \nonumber
 \end{eqnarray}
where ${\bf j}=(c/4\pi)\nabla\times{\bf B}$.

\section{Current sheet examples}\label{sec:cs}
\begin{figure*}
\centering
\includegraphics[width=1\textwidth]{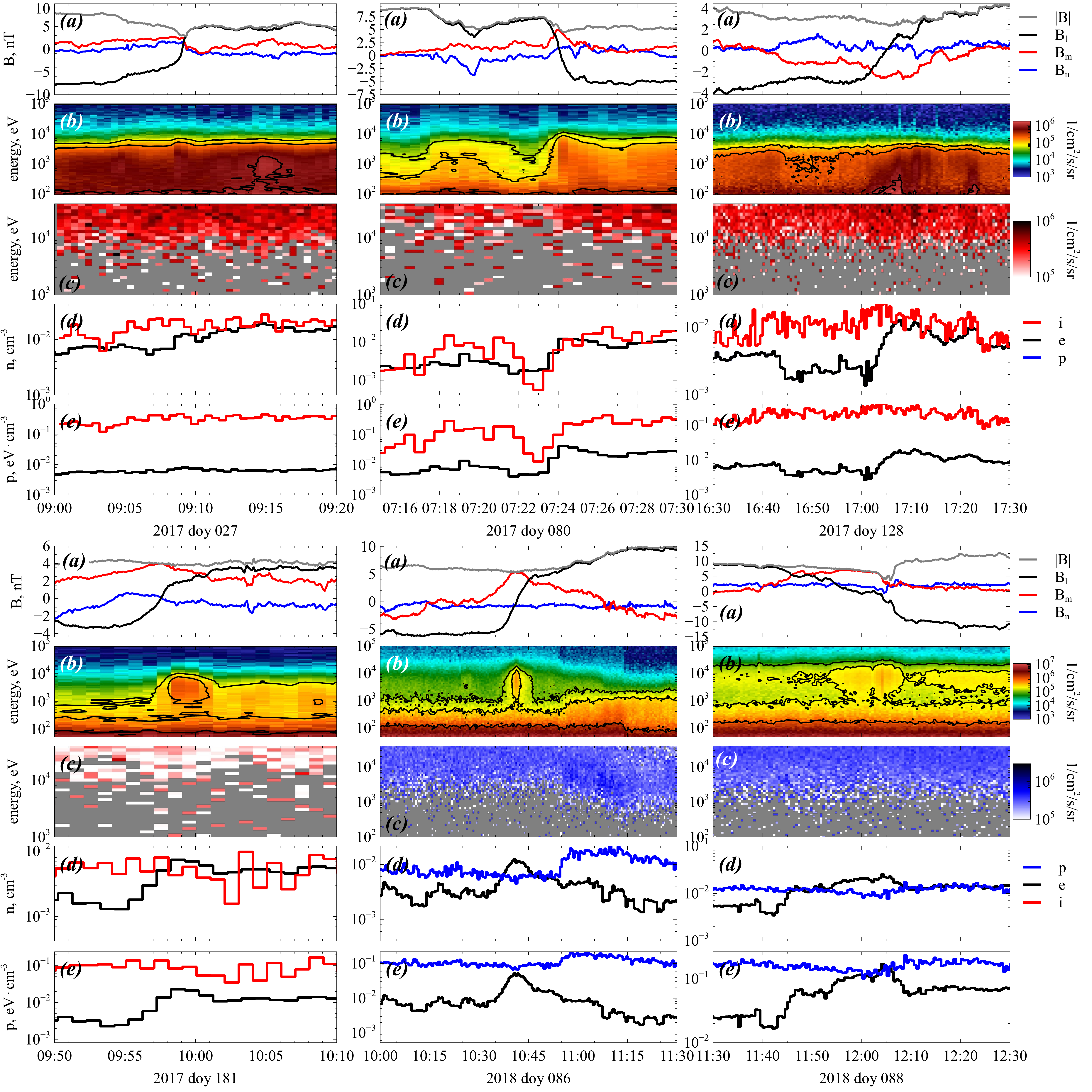}
\caption{Six examples of force-free current sheets observed by Juno in different radial distances (see table \ref{table}). (a) Magnetic field components in the local (MVA) coordinate system and the magnetic field magnitude. (b, c) Omnidirectional spectra of electrons and dominant ion species (blue for protons and red for heavy ions). (d,e) Densities and pressures of electrons and dominant ion species.}
\label{fig3}
\end{figure*}

Figure \ref{fig3} shows six typical examples of thin current sheets with an almost constant magnetic field magnitude across the sheet, $|{\bf B}|\approx const$. Such a constant magnetic field implies the dominant role of field-aligned currents in the current sheet configuration: if ${\bf j}=C\cdot {\bf B}$, then ${\bf j}\times {\bf B}=0$ and there is no pressure variation across the sheet \cite<note that typical crossings of the magnetodisk current sheet show strong variations of the plasma pressure (density) across the sheet, see, e.g.,>{Bagenal21:Huscher_plasmasheet}. Panels (a) show that $|{\bf B}|=const$ is due to peak of $B_m$ component that compensates the drop of $B_l^2$ around the neutral plane (where $B_l=0$). As expected for the force-free current sheet, there are no appreciable variations in the ion fluxes (protons or heavy ions) across the sheet (see panel (c)). Electron fluxes may show some variations (see panel (b)), but variations of the electron thermal pressure are insufficient to cause any significant variations of the magnetic field pressure (see panel (d); note that a variation of $10^{-2}\cdot{\rm cm}^{-3}{\rm keV}$ corresponds to $\approx 2$nT variation of the magnetic field).

Let us explain the absence of ion pressure variations during the observed current sheet crossings. As shown in Fig. \ref{fig3}, the time-scale of current sheet crossings varies from $T \sim 5$ min to $30$ min; taking into account the flapping speed of $\sim \omega_JR\tan\theta$, this time-scale can be converted to a spatial scale $L\approx 1000{\rm km}\cdot (r/50R_J)\cdot (T/60{\rm s}) \in[5,30]\cdot 10^3km$ (here $\omega_J$ is the Jupiter rotational frequency, $r$ is the radial distance of the current sheet, and $\theta\approx 9.5^\circ$ is the tilt angle of the magnetodisk with respect to the planetary equator, see \citeA{Connerney98,Khurana05}). Despite that we used the upper limit for the flapping speed \cite<see discussion and observations in>{Hill79,Kim20:juno:PS},
this spatial scale is much smaller than the typical thicknesses of Jovian magnetodisk current sheets, $L\sim 2R_J\approx 10^5$km \cite{Connerney20:Magnetodisc_Model,Liu21:juno,Khurana22}. More importantly, this spatial scale is comparable to (or smaller than) the hot proton or heavy ion gyroradius: for the equatorial field of a typical current sheet, $\sim 5$nT, $\sim 30$keV protons and sulfur ions have gyroradii of $\sim 5000$km and $\sim 25000$km, respectively. Thus, these current sheets are likely on sub-ion scale, within which ions cannot redistribute their pressure. To establish the pressure balance in such current sheets, the field-aligned electron currents create a local $B_m$ peak.

\begin{figure*}
\centering
\includegraphics[width=1\textwidth]{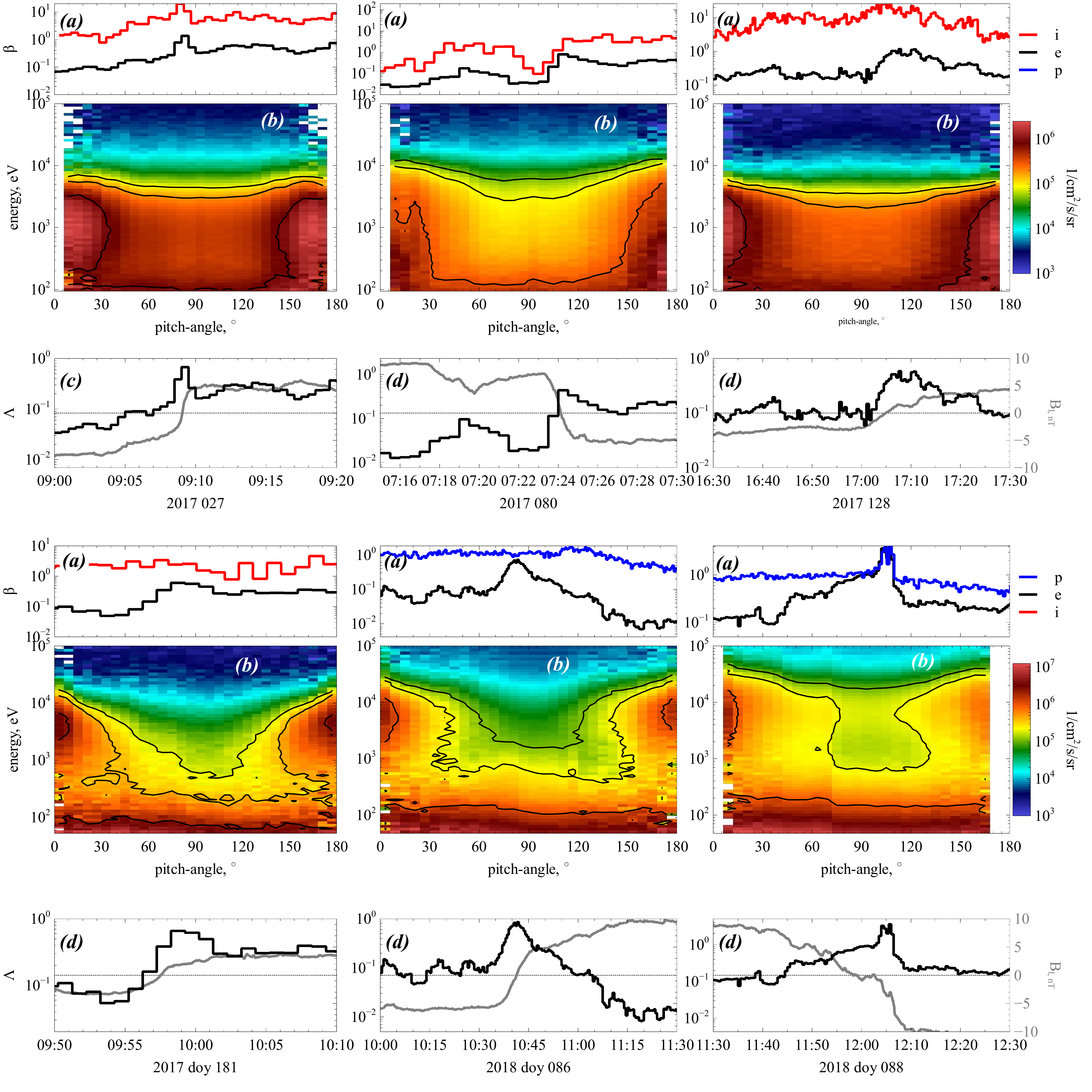}
\caption{Six examples of force-free current sheets from Fig. \ref{fig3}. (a) Electron and ion betas. (b) Electron pitch-angle, energy distribution averaged over the entire event. (c) Electron fire-hose parameter and $B_l$ field to illustrate the current sheet center.}
\label{fig3b}
\end{figure*}

To estimate the electron contribution to the field-aligned currents, we use Eq. (\ref{eq:currents_par}) and the measured electron pitch-angle, energy distributions. Figure \ref{fig3b}(b) shows that all six current sheets are characterized by field-aligned bi-directional electron streams. These streams occupy $\sim30^\circ$ in pitch angles around the parallel and anti-parallel (with respect to the background magnetic field) directions, in the energy range below $\sim 10$ keV. Such field-aligned streams may be generated by reconnection further downtail \cite<e.g.,>{Kronberg12:icarus} or originate from the aurora acceleration region \cite{Mauk17,Mauk20,Elliott20,Allegrini20}. In the Earth's magnetosphere, similar field-aligned streams are observed in the near-Earth magnetotail \cite{Hada81,Walsh13,Artemyev17:grl:currents}, but their energies are well below $\sim 1$keV, in agreement with the capability of the Earth’s aurora acceleration \cite{Ergun04,Chaston07,Watt&Rankin09:prl}. More effective aurora acceleration in the Jupiter magnetosphere may produce $\sim 10$ keV beams \cite{Kollmann18,Saur18,Damiano19:jupiter,Lysak21}, which are likely further expanded in the pitch-angle space by various scattering mechanisms and form the electron streams observed in the plasma sheet \cite<see discussion in>{Zhang20:jupiter}. In the presence of a large electron $\beta_e\sim 1$, such field-aligned streams create a strong pressure anisotropy with the fire-hose parameter $\Lambda_e$ reaching one (see Fig. \ref{fig3b}(c)). Thus, Eq. (\ref{eq:currents_par}) shows that for this large $\Lambda_e$, electrons will carry almost all the current to support $B_l$ and $B_m$ variations across the sheet.

\begin{figure*}
\centering
\includegraphics[width=1\textwidth]{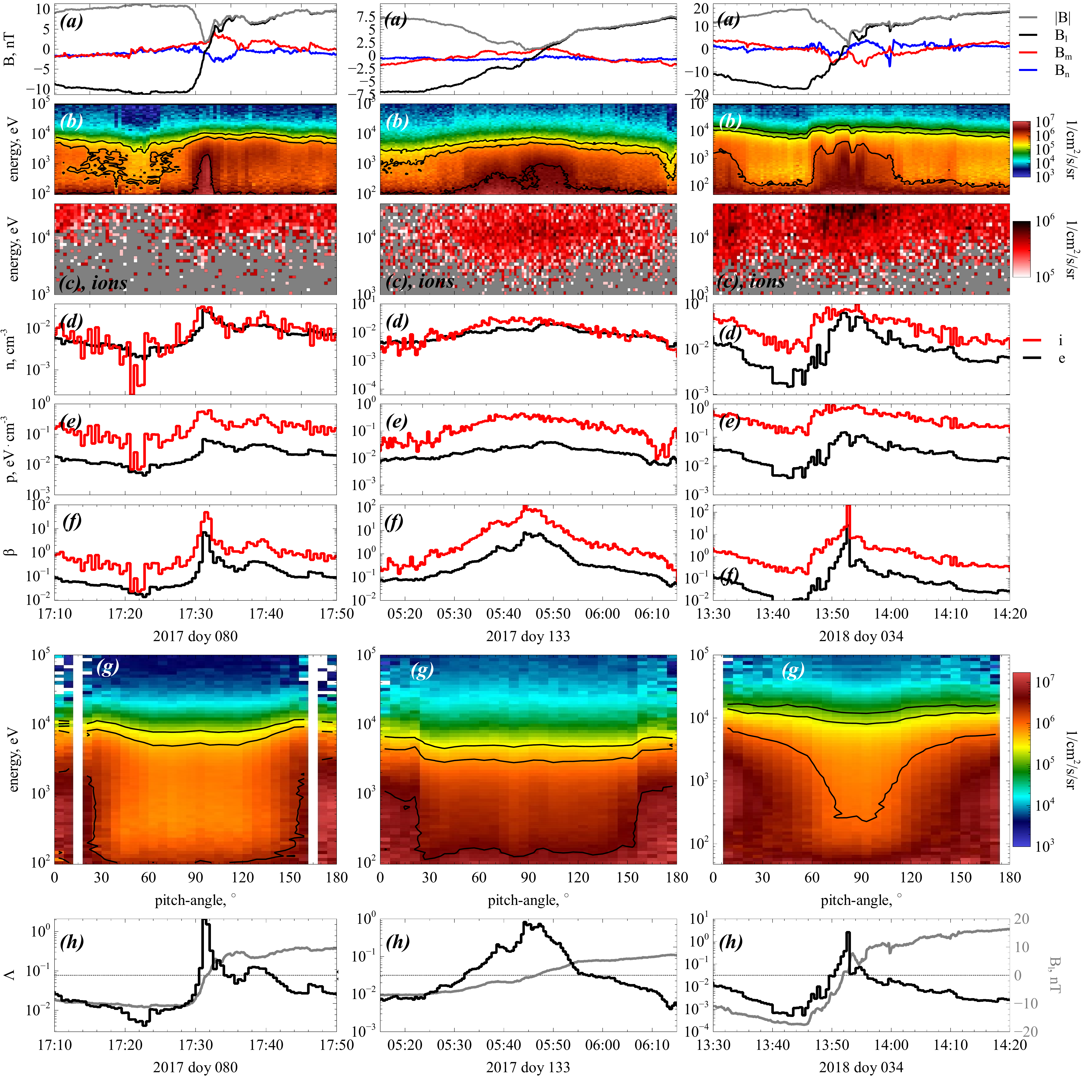}
\caption{Three examples of current sheets with large field-aligned electron currents and plasma pressure variations observed by Juno in different radial distances (see table \ref{table}). (a) Magnetic field components in the local (MVA) coordinate system and the magnetic field magnitude. (b, c) Omnidirectional spectra of electrons and dominant ion species. (d, e) Densities and pressures of electrons and dominant ion species. (f) Electron and ion beta. (g) Electron pitch-angle and energy distribution averaged over the entire event. (h) Electron fire-hose parameter and $B_l$ field to show the current sheet center.}
\label{fig4}
\end{figure*}

We explain the formation of force-free (partial force-free) current sheets with strong field-aligned electron currents (shown in Fig. \ref{fig3b}) as a need to balance the magnetic field pressure decrease $\sim B_l^2$ on a sub-ion scale. This explanation implies that similar electron currents should be observed in current sheets on larger scales, where they will not create $B_m$ peaks, but rather contribute to the cross-field current density, in agreement with Eq. (\ref{eq:currents_per}) \cite<see discussion in>{Artemyev16:pop:cs}. Figure \ref{fig4} shows such current sheets with a significant ion pressure contribution to the stress balance and strong field-aligned electron anisotropy. There is almost no $B_m$ peak in the current sheet center (where $B_l\sim 0$, see panel (a)), whereas ion fluxes and pressures exhibit peaks (panels (c,d)). Temporal scales of current sheet crossings from Fig. \ref{fig4} are about 20-30min, corresponding to a spatial scale larger than the typical proton and heavy ion gyroradius in these current sheets. Note the two current sheets on 2017 doy 080 (shown in Fig. \ref{fig3} and \ref{fig4}) exhibit different characteristics: the force-free current sheet in Fig. \ref{fig3} was crossed within a couple of minutes and shows no $p_i$ variations, whereas the one in Fig. \ref{fig4} was crossed within $\sim 15$ min and shows strong $p_i$ variations.

In current sheets from Fig. \ref{fig4}, electron pitch-angle distributions contain strong field-aligned streams with characteristics very similar to those in the force-free current sheets (compare Figs. \ref{fig3b}(b) and \ref{fig4}(g)). However, contrary to force-free current sheets, these field-aligned streams mostly contribute to anisotropic cross-field electron currents, see Eq. (\ref{eq:currents_per}). Indeed, small magnetic field pressure at the current sheet center (where $B_l\sim 0$) increases $\beta_e$ and leads to a large electron fire-hose parameter $\Lambda_e\approx 1$ even for a moderate anisotropy $A_e$ (see Fig. \ref{fig4}(f,h)).

Comparison of Figs. \ref{fig3}, \ref{fig3b}, and \ref{fig4} suggests the following mechanism for the formation of the force-free current sheet. Certain external conditions (e.g., electron acceleration in the aurora region, see \citeA{Mauk17,Saur18,Damiano19:jupiter,Lysak21}) generate field-aligned electron streams bouncing within the current sheet in the Jovian magnetodisk. These streams contribute to the field-aligned electron anisotropy, $A_e>1$, and fire-hose parameter $\Lambda_e>0$. In typical thick current sheets such anisotropy supports the cross-field electron currents (see Eq.(\ref{eq:currents_per})) and may create a thin, sub-ion scale current sheet embedded into a thick, ion scale current sheet \cite<e.g.,>{Zelenyi04,Mingalev18,Kamaletdinov20:pop,Zelenyi22}. Indeed, the magnetic field profiles in Fig. \ref{fig4} exhibit stronger gradients around $B_l\sim 0$. If external drivers result in the current sheet thinning, the current sheet may reach sub-ion spatial scale where ions cannot redistribute their pressure and maintain the stress balance. In this case, the electron currents form $B_m$ peaks to balance the $B_l^2$ drop at the current sheet center, and self-consistently evolve from the cross-field currents to field-aligned currents (see Schematic in Fig. \ref{fig1}).

\section{Discussion}\label{sec:discussion}
In this study, we investigate force-free (and partially force-free) current sheets, where field-aligned electron streams support the pressure anisotropy and parallel currents, leading to the formation of the $B_m$ peak at the current sheet center, $B_l=0$. Let us discuss the difference of the stress balance in such current sheets from that in more typical {\it thick} current sheets. In current sheets, the 2D stress balance in the equatorial plane (balance along the radial direction) is maintained by a combination of the centrifugal force, plasma pressure force, and magnetic field line tension force \cite{Hill&Carbary78,Cheng83,Zimbardo89}:
\begin{equation}
\frac{1}{c}j_\varphi  B_\theta   + m_i n\omega _J^2 r + \nabla _r \hat p = 0
\end{equation}
where $\nabla_r\hat p$ is the radial component of the plasma pressure tensor gradient. In the local coordinate system, ${\bf l}\approx {\bf e}_r$ and ${\bf m}\approx {\bf e}_\phi$. Based on the vertical stress balance, $8\pi p = \max B_l^2  - \max B_m^2$, we may estimate the current density as:
\begin{equation}
j_m  \approx \frac{c}{{4\pi }}\frac{{\max B_l }}{{B_n  }}\nabla _r \max B_l  \approx 5.5\frac{{{\rm nA}}}{{{\rm m}^{\rm 2} }} \cdot \left( {r/R_J } \right)^{ - 2}
 \approx 12\frac{{{\rm pA}}}{{{\rm m}^{\rm 2} }} \cdot \left( {\frac{r}{{30R_J }}} \right)^{ - 2}
\end{equation}
where $\max B_l\approx 50{\rm nT}\cdot(r/R_J)^{-1}$ and $\max B_l/B_n \approx 20$ are the empirical relations \cite<see, e.g.,>{Artemyev14:pss,Liu21:juno}.
The corresponding current sheet thickness $L= c\max B_l/4\pi j_m \approx 1.5R_J\cdot (r/30R_J)^{-1}$ should be larger than $1R_J$ at $r>30R_J$, which is consistent with the thickness estimates for typical {\it thick} current sheets \cite<e.g.,>{Khurana04,Liu21:juno}. Such current sheets will be crossed during an interval of $\Delta t=L/\omega_j r\tan\theta\approx 3{\rm hours}\cdot (r/30R_J)^{-2}$; for much thinner current sheets as in our dataset, the traversal timescale will be less than $10$ min (see Fig. \ref{fig3}). The stress balance in such thin current sheets cannot be maintained by centrifugal force and radial gradient of the plasma pressure. Instead, the electron pressure anisotropy contributes to the stress balance \cite{Rich72}:
\begin{equation}
\nabla _r \hat p = \nabla _r p_ \bot   + \nabla _\theta \frac{{p_{\parallel e}  - p_ {\bot e}  }}{{B^2 }}B_r B_\theta   \approx \nabla _\theta \left(\frac{\Lambda_e}{4\pi}B_r B_\theta \right)
\end{equation}
This equation shows that the thin current sheet configuration resembles a classical rotational discontinuity, with no variations of the Alfven speed because of the pressure anisotropy \cite{Hudson70}:
\begin{equation}
\Delta v_A=\sqrt{\frac{B^2}{4\pi n m_i}\left(1-\Lambda_e\right)}\sim 0
\end{equation}
This condition allows for a balance of the 1D current sheet (with thickness $L$ much smaller than the spatial scale of the radial gradient of the plasma density) without fast plasma flows typical for rotational discontinuities in anisotropic plasma, where cross-sheet change of the plasma flow velocity equals to $\Delta v_A$ \cite<see discussions of the anisotropy contribution to the force-free current sheet configurations in>{Vasko14:angeo_by,Artemyev20:apjl}. Formation of such 1D current sheets around the fire-hose marginally stability threshold has been predicted theoretically \cite<e.g.,>{FP76,Cowley78,Cowley&Pellat79}, but never have been detected under quiet geomagnetic conditions in the Earth's magnetotail \cite<see discussion in>{Sitnov06,Artemyev&Zelenyi13}. Observations of these current sheets in the Jovian magnetotail confirm the theoretical predictions, which can further lead to improved current sheet models \cite<see discussion on development of the next generation of current sheet models in>{Sitnov&Merkin16,Yoon21:NatCom,Zelenyi22}.

It is worth to note that these current sheets are electromagnetically {\it disconnected} from the Jovian ionosphere, because the local Alfven speed is zero, $v_A=(B/\sqrt{4\pi n m_i})\cdot\sqrt{1-\Lambda_e}=0$, and there are no field-aligned perturbations propagating from the current sheet to the ionosphere. Such local destruction of magnetosphere-ionosphere coupling is an interesting phenomenon that we do not observe in the Earth's magnetotail, where $\Lambda_e$ is much more moderate  \cite{Artemyev20:jgr:electrons}.


Theoretical investigations of these force-free current sheets in the Jovian magnetodisk is a real challenge for plasma kinetics, because these current sheets share properties of 2D plasma equilibria (with the tension force $\sim (4\pi/c)\cdot j_mB_n$) and properties of rotational discontinuities (with ${\bf j}=C\cdot {\bf B}$ and $B_n\ne 0$). All existing 2D kinetic current sheet models operate with the plasma pressure gradients, $\nabla p=c^{-1}{\bf j}\times{\bf B}$, and do not include field-aligned currents \cite<see, e.g.,>[and references therein]{YL05,Vasko13:pop,Zelenyi11PPR,Sitnov06,Sitnov&Merkin16}. Existing models of force-free current sheets mostly assume 1D tangential discontinuities with $B_n=0$ \cite<see, e.g.,>[and references therein]{Harrison09,Panov11:magnetopause,Neukirch20,Neukirch20:jpp}. Construction of the kinetic model for 1D rotational discontinuities requires assumptions of an additional system symmetry \cite<e.g.,>{Sonnerup&Su67,Artemyev11:pop,Mingalev12,Vasko14:angeo_by}, whereas kinetic models of 2D rotational discontinuities have not yet been constructed. Although fluid models of 2D rotational discontinuities (with ${\bf j}=C\cdot {\bf B}$, $j_nB_m=j_mB_n$, and $B_n\ne 0$) can be constructed \cite<e.g.,>{Cowley78,Hilmer87,Hau&Voigt92,Lukin18}, their kinetic realization has not been demonstrated. Therefore, further theoretical investigations are needed to explain Juno observations in the Jovian magnetodisk.

\section{Conclusions}
We have investigated thin (thickness is smaller than or about the hot ion gyroradius) current sheets observed by Juno in the Jovian magnetodisk and characterized by strong field-aligned electron streams. We have demonstrated that electron streams support the strong field-aligned anisotropy, which may increase the fire-hose parameter up to the instability threshold, $\Lambda_e=4\pi(p_{e\parallel}-p_{e\perp})/B^2\to 1$. In such thin, marginally stable current sheets, almost all current is field-aligned, and the current sheet configuration is force-free, with ${\bf j}\times{\bf B}\approx 0$. This is a new type of strongly anisotropic, force-free current sheets, which has not been reported in the quiet-time Earth's magnetopshere or solar wind. Further numerical investigations of such current sheet formation and dynamics will reveal their potential role in the particle acceleration (e.g., via magnetic reconnection).

\acknowledgments
This work is supported by Grants 80NSSC19K1593 (A.V.A.) under Juno Participating Scientist program, and by subcontract 699046X to UCLA under prime contract ZZM06AA75C (X.J.Z. and Q.M.).

\section*{Open Research} \noindent
Data processing was done using SPEDAS V4.1, see \citeA{Angelopoulos19}.
\noindent All the adopted data are available in the archive \url{https://doi.org/10.5281/zenodo.7470240}


\end{document}